%% file: LightStopPaper.tex
\RequirePackage{lineno}
\documentclass[epj,nopacs]{svjour}

\usepackage{subfigure}
\usepackage{mathrsfs}
\usepackage{graphicx}
\usepackage{multirow}
\usepackage{rotating}
\usepackage{atlasphysics}
\usepackage{url}
\usepackage{hyperref}
\usepackage{textpos}
\usepackage{color}
\usepackage{cite}
\usepackage{multirow}

\usepackage{preprintcover} 
\PreprintCoverPaperTitle{Search for light scalar top quark pair production in final states with two leptons with the ATLAS detector in $\sqrt{s}=7\TeV$ proton--proton collisions} 
\PreprintIdNumber{CERN-PH-EP-2012-211}  
\PreprintCoverAbstract{A search is presented for the pair production of light scalar top quarks in $\sqrt{s}=7$~TeV proton--proton collisions recorded with the ATLAS detector at the Large Hadron Collider. This analysis uses the full data sample collected during 2011 that corresponds to a total integrated luminosity of 4.7~fb$^{-1}$. Light scalar top quarks are searched for in events with two opposite-sign leptons ($e$, $\mu$), large missing transverse momentum and at least one jet in the final state. No excess over Standard Model expectations is found, and the results are interpreted under the assumption that the light scalar top decays to a $b$-quark in addition to an on-shell chargino whose decay occurs through a virtual $W$ boson. If the chargino mass is 106 GeV, light scalar top quark masses up to 130~GeV are excluded for neutralino masses below 70~GeV. }  
\PreprintJournalName{EPJC}

\input{susydefs}
\usepackage{setspace}
\titlerunning{Search for light scalar top quark pair production in final states with two leptons}\newcommand{\dileptitle}{Search for light scalar top quark pair production in final states with two leptons with the ATLAS detector in $\sqrt{s}=7\TeV$ proton--proton collisions}

\newcommand{\hideit}[1]{{ }}

\newcommand{\alpgen}{{\tt ALPGEN 2.14}}
\newcommand{\powheg}{{\tt POWHEG }}

\newcommand{\herwig}{{\tt HERWIG 6.520}}
\newcommand{\pythia}{{\tt PYTHIA 6.425}}
\newcommand{\mcatnlo}{{\tt MC@NLO 4.01}}
\newcommand{\fewz}{{\tt FEWZ}}
\newcommand{\mcfm}{{\tt MCFM}}

\newcommand{\jimmy}{{\tt JIMMY 4.31}}

\begin{document}

\title{\dileptitle \begin{textblock}{6}[0,1](0,-3){\small CERN-PH-EP-2012-211, for EPJC Letters}
\end{textblock}
}

\author{The ATLAS Collaboration}
\institute{}

\date{}

\abstract{A search is presented for the pair production of light scalar top quarks in $\sqrt{s}=7$~TeV proton--proton collisions recorded with the ATLAS detector at the Large Hadron Collider. This analysis uses the full data sample collected during 2011 running that corresponds to a total integrated luminosity of 4.7~fb$^{-1}$. Light scalar top quarks are searched for in events with two opposite-sign leptons ($e$, $\mu$), large missing transverse momentum and at least one jet in the final state. No excess over Standard Model expectations is found, and the results are interpreted under the assumption that the light scalar top decays to a $b$-quark in addition to an on-shell chargino whose decay occurs through a virtual $W$ boson. If the chargino mass is 106 GeV, light scalar top quark masses up to 130~GeV are excluded for neutralino masses below 70~GeV.}

\maketitle

\section{Introduction}

Weak-scale supersymmetry (SUSY)~\cite{Miyazawa:1966,Ramond:1971gb,Golfand:1971iw,Neveu:1971rx,Neveu:1971iv,Gervais:1971ji,Volkov:1973ix,Wess:1973kz,Wess:1974tw} is an extension to the Standard Model (SM) 
that provides a solution to the instability of the scalar SM sector with 
respect to new high-scale physics. For each known boson or fermion, SUSY 
introduces a particle with identical quantum numbers except for a 
difference of half a unit of spin. 
In the framework of a generic $R$-parity conserving minimal supersymmetric 
extension of the SM (MSSM)~\cite{Fayet:1976et,Fayet:1977yc,Farrar:1978xj,Fayet:1979sa,Dimopoulos:1981zb}, SUSY particles are produced in pairs and 
the lightest supersymmetric particle (LSP) is stable. In a large variety of models, 
the LSP is the lightest neutralino, $\tilde{\chi}^{0}_{1}$, which is only weakly interacting.
The scalar partners of right-handed and left-handed quarks, $\tilde q_{R}$ and $\tilde q_{L}$,  
mix to form two mass eigenstates, $\tilde q_1$ and $\tilde q_2$, with $\tilde q_1$ defined to be the lighter 
one. In the case of the supersymmetric partner of the top quark ($\tilde{t}$, stop), 
large mixing effects can lead to one stop mass eigenstate, $\tilde{t}_1$, that is 
significantly lighter than the other squarks.  
Depending on the SUSY particle mass spectrum, stop pair production and decay can result in final 
states topologically similar to $t\bar{t}$ events. 

In this Letter, a search for direct stop pair production is presented in $\sqrt{s}=7$~TeV proton--proton collisions recorded with the ATLAS detector at the Large Hadron Collider, considering 
a SUSY particle mass hierarchy such that $m_t > m_{\tilde{t}_1} > (m_{\tilde{\chi}^\pm_1} +m_b$) 
and the $\tilde{t}_1$ decays exclusively via $b+\tilde{\chi}^\pm_1$.  
The mass of all other supersymmetric particles are set to be above 2~TeV, 
and large stop gauge mixing results in $m_{\tilde{t}_2} \gg m_{\tilde{t}_1}$ so that only $\tilde{t}_1$ pair production 
is considered. The stop is predominantly right-handed, but this has little effect on the acceptance and efficiency for the final interpretation. The chargino ($\tilde{\chi}^\pm_1$) mass is set to 106~GeV (above the 
present exclusion limit of 103.5 GeV~\cite{lepstop}) and it is assumed to decay through a virtual $W$ boson ($\tilde{\chi}^\pm_1 \rightarrow W^{*} \tilde{\chi}^0_1 $). The choice of chargino mass is identical to that used in a previous study reported by the CDF experiment~\cite{Aaltonen:2009sf}, thus allowing easy comparison of the CDF and ATLAS results.
Stops within a mass range between 110 GeV and 160 GeV would be produced with 
relatively large cross-sections -- between 245~pb and 41~pb. In this search, dilepton final states ($\ell = e, \mu$) are considered. Although these events could contribute to an anomaly in the measured $t\bar{t}$ cross-section, the relative contribution would be small due to the low transverse momenta of the visible decay products. Events are required to contain at least one 
energetic jet, large missing transverse momentum ($E^\mathrm{miss}_\mathrm{T}$) and low transverse momenta ($p_\mathrm{T}$) leptons, to target the light stop final state.

By targeting very light top squarks, this analysis is complementary to other direct stop searches recently presented by the ATLAS experiment~\cite{Collaboration:2012si,Collaboration:2012ar}.

\section{The ATLAS detector}
The ATLAS detector~\cite{Aad:2008zzms} is a multi-purpose particle
physics detector with a forward-backward symmetric cylindrical
geometry and nearly 4$\pi$ coverage in solid angle\footnote{
ATLAS uses a right-handed coordinate system with its origin at the nominal
interaction point in the centre of the detector and the $z$-axis along the
beam pipe. Cylindrical coordinates $(r,\phi)$ are used in the transverse
plane, $\phi$ being the azimuthal angle around the beam pipe. The pseudorapidity $\eta$ is
defined in terms of the polar angle $\theta$ by $\eta=-\ln\tan(\theta/2)$.
}.  It contains four superconducting magnet systems, which comprise a 
solenoid surrounding the inner tracking detector (ID), and the barrel and two end-cap toroids equipping a muon spectrometer. The ID consists of a silicon pixel detector, a silicon microstrip detector (SCT), and a transition
radiation tracker (TRT).  In the pseudorapidity
region $\left|\eta\right| < 3.2$, high-granularity liquid-argon (LAr)
electromagnetic (EM) sampling calorimeters are used.  An iron/scintillator tile
calorimeter provides coverage for hadron detection over
$\left|\eta\right| < 1.7$.  The end-cap and forward regions, spanning
$1.5~<~\left|\eta\right|~<~4.9$, are instrumented with LAr calorimeters
for both EM and hadronic measurements. The muon spectrometer surrounds the calorimeters and consists of a system of precision tracking chambers ($|\eta|<2.7$), and detectors for triggering ($|\eta|<2.4$).

\section{Simulated event samples}
\label{sec:MC}
Monte Carlo (MC) simulated event samples are used to develop and validate the analysis procedure and to evaluate the SM backgrounds in the signal region. Production of top quark pairs is simulated with \mcatnlo~\cite{mcatnlolong}, using a top quark mass of $172.5$~GeV. Samples of $W(\to\ell\nu)$ and $Z/\gamma^*(\to\ell\ell)$, produced with accompanying jets (of both light and heavy flavour), are obtained with \alpgen~\cite{Mangano:2002ea}. Diboson ($WW$, $WZ$, $ZZ$) production is simulated
with \herwig~\cite{herwig} and single top production 
with \mcatnlo. Fragmentation and hadronisation for the \alpgen~and \mcatnlo~samples are performed with \herwig, using
\jimmy~\cite{Butterworth:1996zw} for the underlying event. Expected diboson yields are normalised using next-to-leading-order (NLO) QCD predictions obtained with \mcfm~\cite{Campbell:1999ah,Campbell:2011bn}. The top-quark contribution is normalised to approximate next-to-next-to-leading-order (NNLO) calculations~\cite{Aliev:2010zk}. The inclusive $W$ and $Z/\gamma^*$ production cross-sections are normalised to the NNLO cross-sections obtained using \fewz~\cite{Gavin:2010az}. \alpgen~and \powheg~\cite{powheg}~samples are used to assess the systematic uncertainties associated with the choice of generator for $t\bar{t}$ production, and {AcerMC}~\cite{Kersevan:2004yg} samples are used to assess the uncertainties associated with initial- and final-state radiation (ISR/FSR)~\cite{ATLAS:2012al}. The choice of the parton distribution functions (PDFs) depends on the generator. CT10~\cite{Lai:2010vv} sets are used for all MC@NLO samples. {MRST LO**}~\cite{Sherstnev:2008dm} sets are used with {\tt HERWIG} and {\tt PYTHIA}, and {CTEQ6L1}~\cite{cteq6l} with \alpgen. The stop production models are simulated using \pythia~\cite{pythia}. Signal cross-sections are calculated to next-to-leading order in the strong coupling constant, including the resummation of soft gluon emission at next-to-leading-logarithmic accuracy (NLO+NLL) \cite{Beenakker:1997ut,Beenakker:2010nq,Beenakker:2011fu}. An envelope of cross-section predictions is defined using the 68\% C.L. ranges of the {CTEQ6.6} (including the $\alpha_S$ uncertainty) and {MSTW 2008 NLO}~\cite{Martin:2009iq} PDF sets, together with independent variations of the factorisation and renormalisation scales by factors of two or one half. The nominal cross-section value is taken to be the midpoint of the envelope and the uncertainty assigned is half the full width of the envelope, following the PDF4LHC recommendations~\cite{Botje:2011sn}. All MC samples are produced using a {GEANT4}-based~\cite{geant4} detector simulation~\cite{atlassimulation}. The effect of multiple proton--proton collisions from the same or different bunch crossings is incorporated into the simulation by overlaying additional {\tt PYTHIA} minimum bias events onto hard-scattering events. Simulated events are weighted to match the distribution of the mean number of interactions per bunch crossing observed in data.

\section{Data and event selection}
The analysis uses the full 2011 proton-proton collision data sample. After applying the beam, detector and data-quality requirements, the data sample corresponds to a total integrated luminosity of 4.7~fb$^{-1}$. Events were triggered using a combination of single and double lepton triggers. The single electron triggers vary with the data-taking period, and the tightest of these has an efficiency of $\sim$97\% for electrons with $p_\mathrm{T} >$ 25 GeV. The single muon trigger used for all data-taking periods reaches an efficiency plateau of $\sim$75\% ($\sim$90\%) in the barrel (end-caps) for muons with $p_\mathrm{T} >$ 20 GeV. All efficiencies are quoted with respect to reconstructed leptons, passing the baseline lepton definitions. The double lepton triggers reach similar plateau efficiencies, but at lower $p_\mathrm{T}$ thresholds (greater than 17 GeV for electrons passing the dielectron trigger, and greater than 12 GeV for muons passing the dimuon trigger; for the electron-muon trigger the thresholds are 15 and 10 GeV for electrons and muons respectively). If a lepton has an offline $p_T$ above the single lepton trigger plateau threshold in a given event, the relevant single lepton trigger is used. Double lepton triggers are used for events with no such lepton. An exception to this rule is applied in the $\mu\mu$ channel. In this case when one lepton has $p_\mathrm{T} >$ 20 GeV and the second $p_\mathrm{T} >$ 12 GeV, a logical OR of both triggers is used to recover efficiency.

Jet candidates are reconstructed using the anti-$k_t$ jet clustering algorithm~\cite{Cacciari:2008gp} with a radius parameter of $0.4$. The inputs to this algorithm are three-dimensional energy clusters seeded by calorimeter cells with energy significantly above the noise resulting from the electronics and additional proton--proton interactions (calorimeter clusters). The jet candidate energies are corrected for the effects of calorimeter non-compensation, inhomogeneities and energy loss in material in front of the calorimeter, by using $p_\mathrm{T}$- and $\eta$-dependent calibration factors based on MC simulations and validated with extensive test-beam and collision-data studies \cite{Aad:2011he}. Furthermore, the reconstructed jet is modified such that the jet direction points to the primary vertex, defined as the vertex with the highest summed track $p_\mathrm{T}^2$. Only jet candidates with corrected transverse momenta $p_\mathrm{T} > 20$ \GeV{} and $|\eta|<4.5$ are subsequently retained. Jets likely to have arisen from detector noise or cosmic rays are rejected~\cite{Aad:2011he}. Electron candidates are required to have $p_\mathrm{T} > 10~\GeV$, $|\eta| < 2.47$, and pass the ``medium''  shower shape and track selection criteria of Ref.~\cite{Aad:2011mk}. Muon candidates are reconstructed using either a full muon spectrometer track matched to an ID track, or a muon spectrometer segment matched to an extrapolated ID track~\cite{muons}. They must be reconstructed with sufficient hits in the pixel, SCT and TRT detectors. They are required to have $p_\mathrm{T} > 10$ \GeV{} and $|\eta| < 2.4$.

Following object reconstruction, overlaps between candidate jets and leptons are resolved. Any jet candidate lying within a distance $\mathrm{\Delta} R = \sqrt{(\mathrm{\Delta} \eta)^2 + (\mathrm{\Delta} \phi)^2}=0.2$ of an electron is discarded. Subsequently, any electron or muon candidate remaining within a distance $\mathrm{\Delta} R =0.4$ of any surviving jet candidate is discarded.  

The measurement of the missing transverse momentum $\mathbf{p}_\mathrm{T}^\mathrm{miss}$, and its magnitude $E_\mathrm{T}^\mathrm{miss}$, is based on the transverse momenta of all electrons, muons and jets as described above, and of all calorimeter clusters with $|\eta|<4.5$ not associated to such objects.  

Following overlap removal, electrons are further required to have $p_\mathrm{T}>17$~GeV and to pass the ``tight''~\cite{Aad:2011mk} quality criteria, which places additional requirements on the ratio of calorimetric energy to track momentum, and the fraction of high-threshold hits in the TRT. Electrons are also required to be isolated: the $p_\mathrm{T}$ sum of tracks above 1~GeV within a cone of size $\mathrm{\Delta} R=0.2$ around each electron candidate (excluding the electron candidates themselves) is required to be less than 10\% of the electron $p_\mathrm{T}$. Muons must have $p_\mathrm{T}>12$~GeV and must be isolated: the $p_\mathrm{T}$ sum of tracks within a cone of size $\mathrm{\Delta} R=0.2$ around the muon candidate is required to be less than 1.8~GeV. Jets are subject to the further requirements $p_\mathrm{T}>25$~GeV, $|\eta|<2.5$ and a ``jet vertex fraction''\footnote{The jet vertex fraction quantifies the fraction of track transverse momentum associated to a jet which comes from the primary vertex. The cut removes jets within the tracker acceptance which originated from uncorrelated soft collisions.} higher than 0.75. 

The top background measurement described below uses a $b$-tagging algorithm~\cite{ATLAS-CONF-2011-102}, which exploits the topological structure of weak $b$- and $c$-hadron decays inside a candidate jet to identify jets containing a $b$-hadron decay. The nominal $b$-tagging efficiency, computed from $t\bar{t}$ MC events, is on average 60\%, with a misidentification (mis-tag) rate for light-quark/gluon jets of less than 1\%. To correct small differences in the $b$-tagging efficiency observed in the simulation with respect to the data, a scale factor is applied to all simulated samples.

During part of the data-taking period, a localised electronics failure in the electromagnetic calorimeter created a dead region ($\mathrm{\Delta}\eta \times \mathrm{\Delta}\phi \approx 1.4 \times 0.2$). For jets in this region, a correction to their energy is made using the energy depositions in the neighbouring cells, and is propagated to $E^\mathrm{miss}_\mathrm{T}$. If the energy correction exceeds 10 GeV or 10\% of the $E^\mathrm{miss}_\mathrm{T}$, the event is discarded. Events with reconstructed electrons in the calorimeter dead region are also rejected.

\begin{table*}[!tp]
\begin{center}
\small
\begin{tabular}{|c| c c c|}
\hline
Requirement & $ee$ channel & $\mu\mu$ channel & $e\mu$ channel \\
\hline
\multicolumn{4}{|c|}{Signal Region}\\
\hline
lepton $p_\mathrm{T}$ & $>17~$GeV & $>12$~GeV & $>17(12)$~GeV for $e(\mu)$ \\
leading lepton $p_\mathrm{T}$ & \multicolumn{3}{c|}{$<30$~GeV} \\
$m_{ll}$& \multicolumn{2}{c}{$>20$~GeV and $Z$ veto} & $>20$~GeV\\
jet $p_\mathrm{T}$ & \multicolumn{3}{c|}{$\ge 1$ jet, $p_\mathrm{T}>25$~GeV} \\
$E_\mathrm{T}^\mathrm{miss}$ & \multicolumn{3}{c|}{$>20$~GeV}\\
$E_\mathrm{T}^\mathrm{miss,sig}$ & \multicolumn{3}{c|}{$>7.5$~GeV$^{1/2}$}\\
\hline
\multicolumn{4}{|c|}{Top Control Region}\\
\hline
lepton $p_\mathrm{T}$ & $>17~$GeV & $>12$~GeV & $>17(12)$~GeV for $e(\mu)$ \\
leading lepton $p_\mathrm{T}$ &\multicolumn{3}{c|}{$>30$~GeV}\\
$m_{ll}$& \multicolumn{2}{c}{$>20$~GeV and $Z$ veto}& $>20$~GeV \\
jet $p_\mathrm{T}$ & \multicolumn{3}{c|}{$\ge 2$ ($b$)jets, $p_\mathrm{T}>25$~GeV} \\
$b$-jet $p_\mathrm{T}$ & \multicolumn{3}{c|}{$\ge 1$ $b$ jet, $p_\mathrm{T}>25$~GeV}\\
$E_\mathrm{T}^\mathrm{miss}$ & \multicolumn{3}{c|}{$>20$~GeV}\\
$E_\mathrm{T}^\mathrm{miss,sig}$ & \multicolumn{3}{c|}{$>7.5$~GeV$^{1/2}$}\\
\hline
\multicolumn{4}{|c|}{$Z$ Control Region}\\
\hline
lepton $p_\mathrm{T}$ &  $>17~$GeV & $>12$~GeV & n/a\\
leading lepton $p_\mathrm{T}$ & \multicolumn{2}{c}{$<30$~GeV} & n/a \\
$m_{ll}$& \multicolumn{2}{c}{$>81$ GeV and $<101$ GeV}& n/a \\
jet $p_\mathrm{T}$ &\multicolumn{2}{c}{$\ge 1$ jet, $p_\mathrm{T}>25$~GeV} & n/a \\
$E_\mathrm{T}^\mathrm{miss}$ & \multicolumn{2}{c}{$>20$~GeV} & n/a \\
$E_\mathrm{T}^\mathrm{miss,sig}$ & \multicolumn{2}{c}{$>4.0$~GeV$^{1/2}$} & n/a \\
\hline
   \end{tabular}
    \caption{Signal region, top control region and $Z$ control region requirements in each flavour channel. The $Z$ veto rejects events with $m_{ll} > 81$~GeV and $m_{ll} < 101$~GeV. 
    \label{tab:selection}}
  \end{center}
\end{table*}

Events are subject to the following requirements. The primary vertex in the event must have at least five associated tracks and each event must contain exactly two selected leptons (electrons or muons) of opposite sign. Both of these leptons must additionally satisfy the full list of signal lepton requirements, and the dilepton invariant mass, $m_{ll}$, must be greater than $20$~GeV across all flavour combinations. In addition, events in the signal region must have at least one jet with $p_\mathrm{T}>25$~GeV, $E_\mathrm{T}^\mathrm{miss}>20$~GeV, missing transverse momentum significance\footnote{In this paper, $E^\mathrm{miss,sig}_\mathrm{T}=E_\mathrm{T}^\mathrm{miss}/\sqrt{H_\mathrm{T}}$, where $H_\mathrm{T}$ is the scalar sum of the jet and lepton transverse momenta in each event.} $E^\mathrm{miss,sig}_\mathrm{T}>7.5$~GeV$^{1/2}$ to reject multijet events, and leading lepton $p_\mathrm{T}<30$ GeV (to provide further rejection of the dominant dileptonic $t\bar{t}$ background). Events in the $ee$ and $\mu\mu$ channels are subject to a further requirement on the dilepton invariant mass to reject events arising from $Z$ production and decay. This selection, summarised in Table~\ref{tab:selection}, has a low signal efficiency, but strong background rejection. The main factor in the efficiency loss is the lowest lepton $p_\mathrm{T}$ requirement needed to reach the efficiency plateau of the dilepton triggers. The kinematic acceptance varies between $0.06\%$ and $0.3\%$ for a neutralino mass of 55~GeV as the stop mass varies between 112~GeV and 180~GeV, and between $0.1\%$ and $0.003\%$ for a stop mass of 140~GeV as the neutralino mass varies between 1~GeV and 95~GeV (the detector efficiency for these points, defined as the efficiency for reconstructing events that already enter the kinematic acceptance, is $\sim40\%$).

\section{Background estimation}
The dominant SM background,  after the signal selection requirements, arises from $t\bar{t}$ events where both top quarks decay leptonically, with the next most significant background being $Z/\gamma^*+$jets. Single top, $W+$jets, diboson and multijet events give much smaller expected contributions.

The fully leptonic $t\bar{t}$ background in the signal region is obtained by extrapolating the number of $t\bar{t}$ events measured in a suitable control region (CR), after correcting for contamination from non-$t\bar{t}$ events, into the signal region (SR). This extrapolation, detailed in Eq.~\ref{eqn:top}, uses the ratio of the number of simulated $t\bar{t}$ events in the signal region to those in the control region:

\begin{equation}
\small
(N_{t\bar{t}})_\textup{\tiny{SR}}=[ (N_\textup{\tiny{data}})_\textup{\tiny{CR}}-(N_{\textup{\tiny{non}-}t\bar{t},\textup{\tiny{MC}}})_\textup{\tiny{CR}}] \frac{(N_{t\bar{t},\textup{\tiny{MC}}})_\textup{\tiny{SR}}}{(N_{t\bar{t},\textup{\tiny{MC}}})_\textup{\tiny{CR}}}
\label{eqn:top}
\end{equation}

The CR is designed to give an event sample dominated by top events, whilst minimising signal contamination. It is further chosen to be kinematically similar to the signal region to minimise systematic uncertainties due to extrapolation. Selection requirements for the top control region are summarised in Table~\ref{tab:selection}. In this analysis, models with small stop--chargino mass difference are considered, and hence soft $b$-jets are expected in the signal events which are not efficiently tagged. By requiring a $b$-jet in the top control region a high-purity sample of top events is obtained. The signal contamination in the considered models is typically of the order of a few per cent, rising to ~30$\%$ for models with $m_{\tilde{\chi}^0_1}=1$~GeV and high $m_{\tilde{t}_1}$. The percentage of SM, non-$t\bar{t}$ events in the CR is less than $5\%$ across all channels. The resulting $t\bar{t}$ background contributions are consistent with the expected MC yields in all channels within the uncertainties. Signal contamination is taken into account when setting the exclusion limit in the next section by including, for each signal model, the expected signal yield in the top control region in the $(N_{\textup{\tiny{non}-}t\bar{t},\textup{\tiny{MC}}})_\textup{\tiny{CR}}$ term in Eq.~\ref{eqn:top}.

The contribution from $Z/\gamma^*+$jets events to the signal region (from $ee$ and $\mu\mu$ events) is evaluated in a similar way. Data are used to obtain the normalisation of the $Z/\gamma^*$ background in a suitable CR and MC is used to extrapolate from CR to SR using an equation analogous to Eq.~\ref{eqn:top}. This method is used separately for each of the $ee$ and $\mu\mu$ channels (with selection requirements for the $Z$ CR as summarised in Table~\ref{tab:selection}), whereas the contribution to $e\mu$ (including those from $Z/\gamma^*\rightarrow\tau\tau$) is taken directly from the MC simulation due to the limited number of events in the CR. The contamination from non-$Z/\gamma^*+$jets SM events in the CR is less than 5\%, and the signal contamination less than 4\%. The resulting $Z$ background contributions are consistent with the expected MC yields in the $ee$ and $\mu\mu$ channels within the uncertainties. The effect of signal contamination of the $Z$ control region on the final exclusion limit can be neglected to a very good approximation.

Single top, $W+$jets (including heavy-flavour contributions) and diboson backgrounds are evaluated in the signal region directly from the MC simulation. The estimated contribution from $W+$jets has been cross-checked using a data-driven technique (an extension of the ``template fit'', described below), and found to be in good agreement.

\begin{table*}[!tbp]
\begin{center}
\begin{tabular}{|c|c|c|c|c|}
\hline
 & $ee$ & $e\mu$ & $\mu\mu$ & all \\ 
\hline
$t\bar{t}$&$44 \pm 4\pm 5$ & $139 \pm7 \pm 22$& $111\pm 8\pm 10 $ & 293 $\pm$ 12 $\pm$ 34\\
$Z/\gamma^{*}+$jets&$5\pm 1\pm2$ & $23\pm2\pm8$&$48\pm16\pm27$ & 76 $\pm$ 16 $\pm$ 27\\
Single top & $3\pm0.5\pm1$&$12\pm1\pm2$&$12\pm1\pm2$ & 28 $\pm$ 2 $\pm$ 5\\
$W$+jets&$3\pm3\pm3$&$5\pm2\pm1$&$6\pm2\pm1$ & 13 $\pm$ 3 $\pm$ 3\\
Diboson&$4\pm0.4\pm0.5$&$9\pm0.7\pm2$&$10\pm0.7\pm1$ & 22 $\pm$ 1 $\pm$ 3\\
Multijet&2.9 $^{+3.2}_{-2.9}$  $\pm$ 2.2 &2.0 $\pm$ 1.4  $\pm$ 0.3 &3.0 $\pm$ 2.8  $\pm$ 0.3 & 8.0 $\pm$ 3.7 $\pm$ 2.3\\
\hline
Total&$61 \pm6\pm 6$&$189\pm8\pm21$&$190\pm19\pm31$ & 440 $\pm$ 21 $\pm$ 43\\
Data & 48 & 188 & 195 & 431\\
\hline
$\sigma_{\rm vis}$ (exp. limit)  [fb]& 4.9 & 11.1 & 16.2 & 22.0\\
$\sigma_{\rm vis}$ (obs. limit) [fb]& 3.3& 10.9 & 16.9 & 21.0 \\
\hline 
($m_{\tilde{t}_1}$,$m_{\tilde{\chi}^0_1}$)$=(112,55)$~GeV & $44.1\pm4.8$ & $137\pm8$ & $140\pm8$ & $322\pm13$\\
($m_{\tilde{t}_1}$,$m_{\tilde{\chi}^0_1}$)$=(160,55)$~GeV & $8.8\pm1.5$ & $31.4\pm2.7$ & $36.5\pm2.9$ & $76.6\pm4.3$  \\
\hline
\end{tabular}
\caption[]{The expected and observed numbers of events in the signal region for each flavour channel. In the combined flavour column (``all''), the statistical uncertainty (first uncertainty quoted, includes the MC and data statistical errors) on the various background estimates have each been added in quadrature whilst the systematic uncertainties (second uncertainty quoted) have been combined taking into account the correlations between background sources. Observed and expected upper limits at 95\% confidence level on the visible cross-section $\sigma_{\rm vis} = \sigma \times A \times \epsilon$ are also shown.

The expected signal yields and statistical uncertainties on the yields are quoted for the two mass points illustrated in the figures.}
\label{tab:results}
\end{center}
\end{table*}

\begin{figure*}[!tbp]
\begin{center}
\subfigure[]{\includegraphics[scale=0.34]{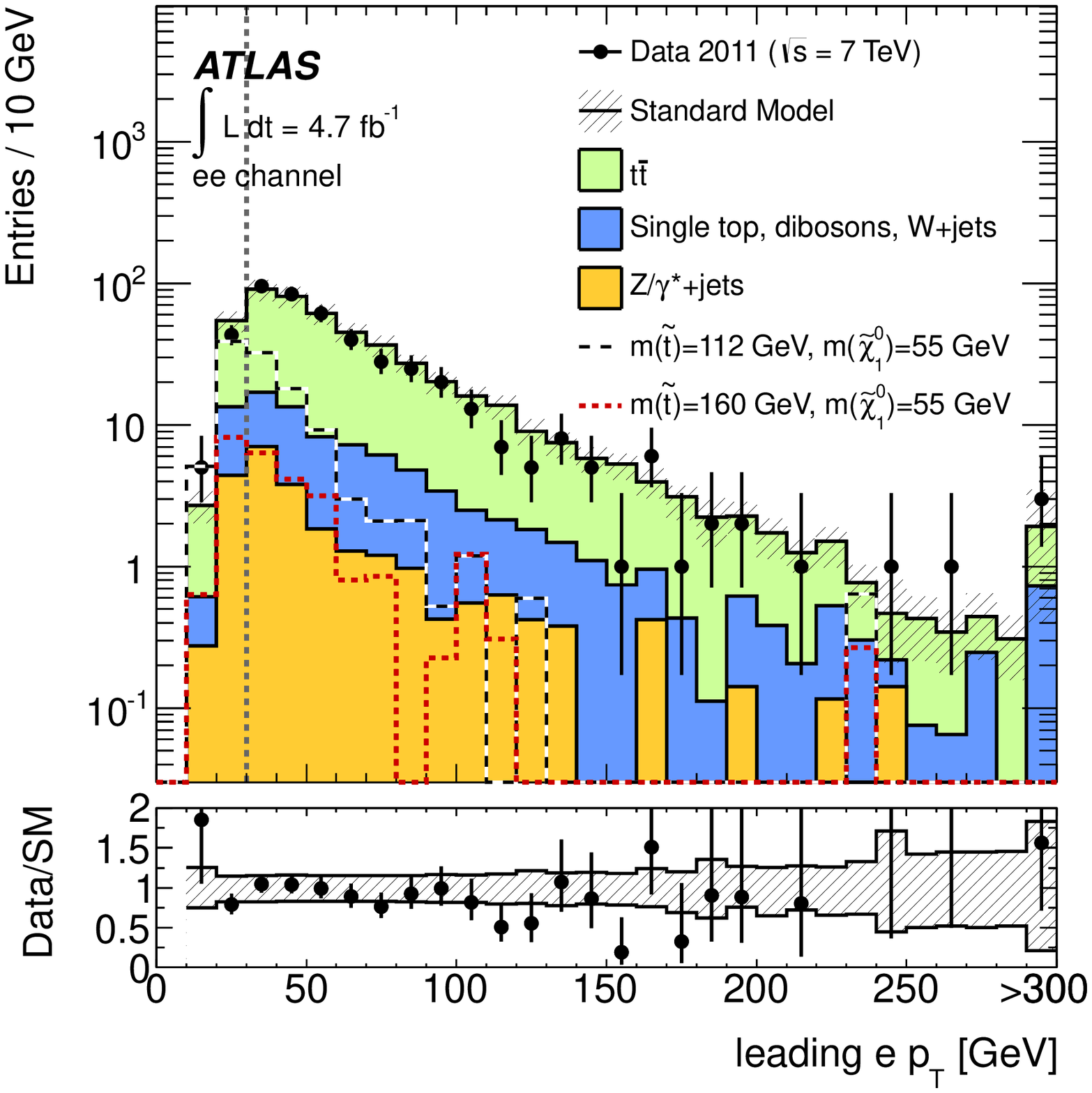}}
\subfigure[]{\includegraphics[scale=0.34]{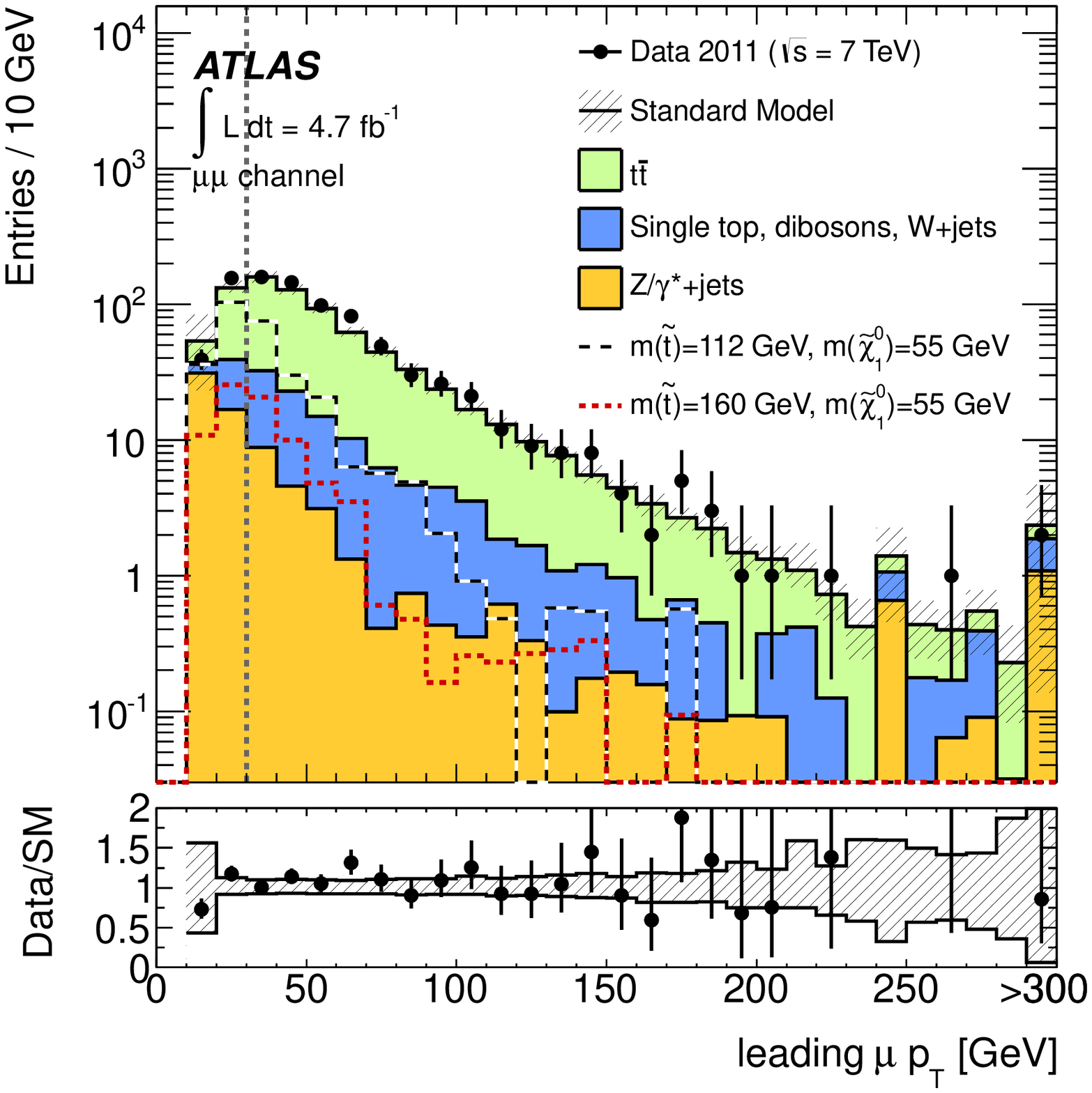}}
\subfigure[]{\includegraphics[scale=0.34]{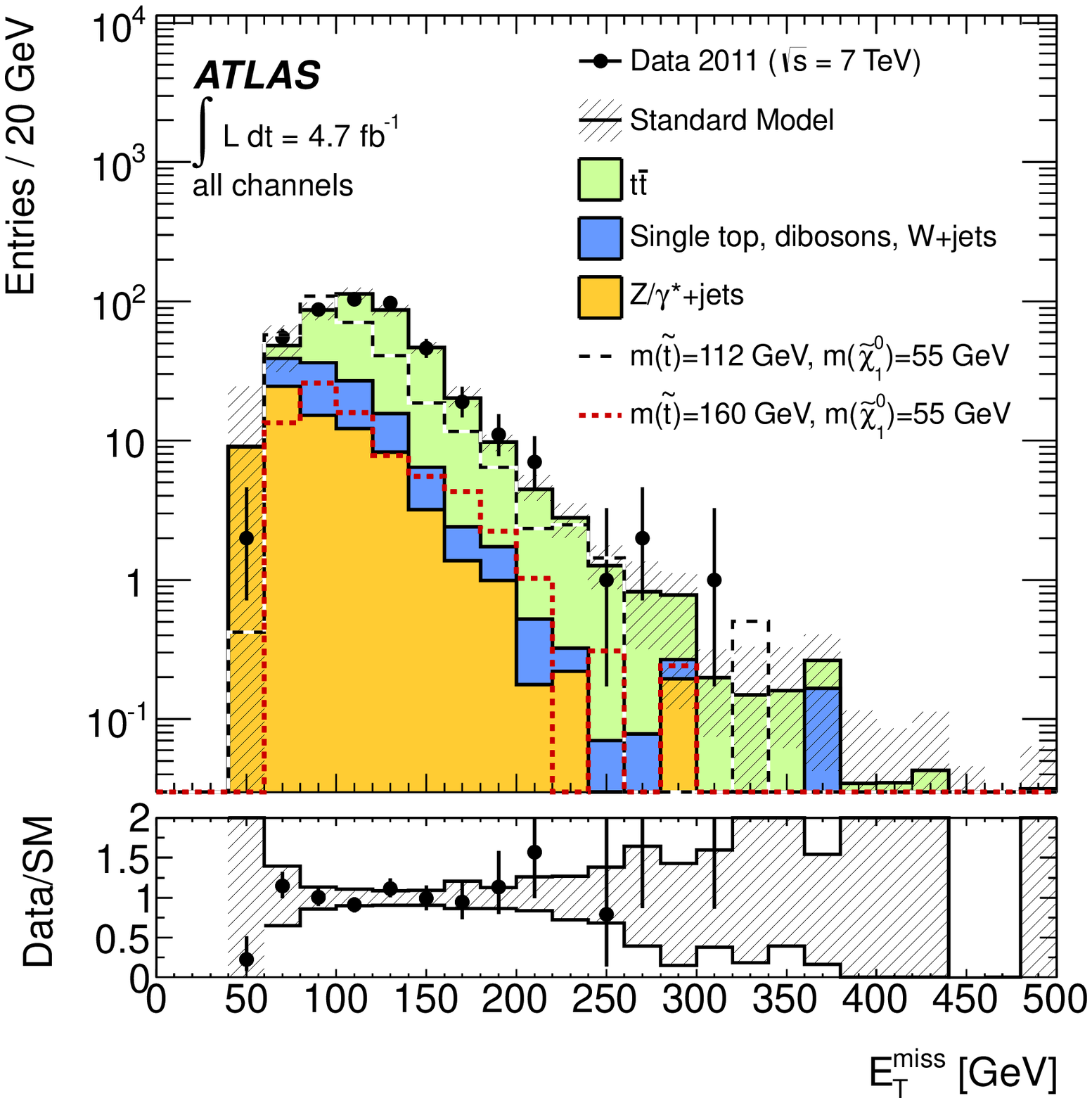}}
\subfigure[]{\includegraphics[scale=0.34]{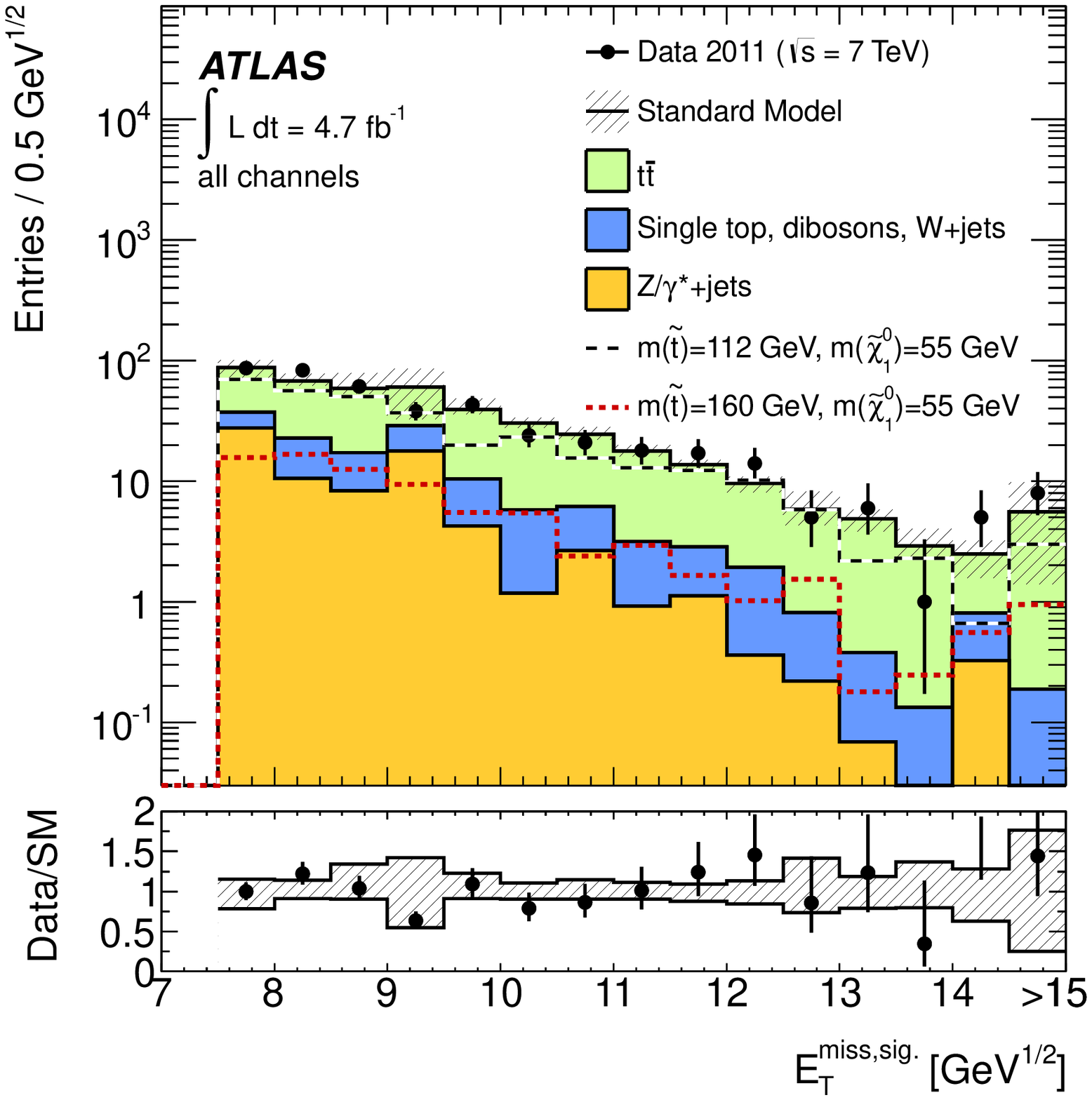}}
\caption{The leading electron and muon $p_\mathrm{T}$ distributions in the same flavour (a)~$ee$ and (b)~$\mu\mu$ channels, before the requirement on the leading lepton $p_\mathrm{T}$ (which is marked by the dashed vertical line), and~(c) the $E^\mathrm{miss}_\mathrm{T}$ distribution and~(d)$E^\mathrm{miss,sig}_\mathrm{T}$ distribution after all signal region requirements. The data and evaluated background components are shown. The hashed band indicates the total experimental uncertainty on the expectation. The dashed lines give the expectations for signal models with stop masses of 112 GeV and 160 GeV, and a neutralino mass of 55 GeV. The last histogram bins in (a) and (b) include the integrals of all events with $p_\mathrm{T}>300$~GeV. The final bin in~(d) includes all events with an $E^\mathrm{miss,sig}_\mathrm{T}$ of at least 15~GeV$^{1/2}$. The bottom panels show the ratio of the data to the expected background (points) and the systematic uncertainty on the background (hashed area). 
\label{leppt_distributions}
}
\end{center}
\end{figure*}

The tight requirement on $E^\mathrm{miss,sig}_\mathrm{T}$ heavily suppresses the multijet background. A data-driven template fit technique is used to verify that this background is small, and to assign an uncertainty on the yield in the signal region. The isolation requirements on the electrons and muons are reversed to enhance the multijet content of selected events. The requirements are inverted in the signal region, prior to application of the $E^\mathrm{miss,sig}_\mathrm{T}$ requirement. The shape of the $E^\mathrm{miss,sig}_\mathrm{T}$ distribution in data for this inverted selection (after subtracting the dominantly electroweak background using the MC simulation) is then compared to the equivalent distribution in data for the ``normal'' isolation requirements in order to validate that inverting the lepton isolation does not distort the shape of the distribution. The ``normal'' and ``inverted'' shapes were found to agree very closely for the full range of distributions considered in the analysis. The inverted  $E^\mathrm{miss,sig}_\mathrm{T}$ distribution is then renormalised to match the distribution after nominal isolation requirements. Passing this correctly normalised template through the remaining requirements gives the multijet yield in the signal region. It is found to be small in all channels, making up less than 2\% of the total background. 

\section{Systematic uncertainties}

The total systematic uncertainty on the expected background in the combined flavour channel (the sum of $ee$, $e\mu$ and $\mu\mu$ events) is $9.8\%$, and is dominated by the uncertainties on the two largest backgrounds (dileptonic $t\bar{t}$ and $Z$+jet events). The largest source of systematic uncertainty on the $t\bar{t}$~background evaluation is the uncertainty on the jet energy scale (JES), with smaller contributions coming from the jet energy resolution (JER) uncertainty~\cite{Aad:2011he}, the theory and MC modelling uncertainties (using the prescriptions described in Ref.~\cite{ISRFSR}), the systematic uncertainties on the $b$-tagging efficiency~\cite{ATLAS-CONF-2011-102}, and the uncertainty arising from the limited numbers of MC and data events. Uncertainties~\cite{Aad:2011mk,muonefficiency,more_unc} in lepton reconstruction and identification (momentum and energy scales, resolutions and efficiencies) give smaller contributions. 

The primary source of uncertainty on the $Z/\gamma^*+$jets background estimate in the combined flavour channel is the jet energy resolution uncertainty, with smaller contributions coming from the statistical and jet energy scale uncertainties. Theoretical uncertainties on the $Z/\gamma^*+$jets background are investigated by varying the PDF and renormalisation scales. An uncertainty on the luminosity of 3.9$\%$~\cite{lumi1,lumi2} is included in the systematic uncertainty calculation for backgrounds taken directly from the MC simulation. The dominant uncertainties on these backgrounds are the jet energy scale and statistical uncertainties. The systematic uncertainty on the multijet yield is obtained by varying the range in which the template fit is performed, and using the maximum deviation of the final yield to assign the uncertainty.

In the considered $m_{\tilde{\chi}^0_1}$--$m_{\tilde{t}_1}$ mass plane the theoretical uncertainty on each of the signal cross-sections is approximately 16$\%$. These arise from considering the  cross-section envelope defined using the 68\% C.L. ranges of the {CTEQ6.6} and {MSTW 2008 NLO} PDF sets, and independent variations of the factorisation and renormalisation scales (see Section~\ref{sec:MC}).
Further uncertainties on the numbers of predicted signal events arise from the JES uncertainty (7--15$\%$), the JER uncertainty (1--7$\%$), the luminosity uncertainty (3.9$\%$), the uncertainties on calorimeter energy clusters used to calculate $E^\mathrm{miss}_\mathrm{T}$ (2--6\%), the statistical uncertainty from finite MC event samples (4--20\%) and smaller contributions from uncertainties on lepton reconstruction and identification, where the quoted ranges display the maximum variation observed using all signal models considered in this analysis.

\section{Results and interpretation}

\begin{figure}[!tbp]
\centerline{
\includegraphics[width=0.5\textwidth]{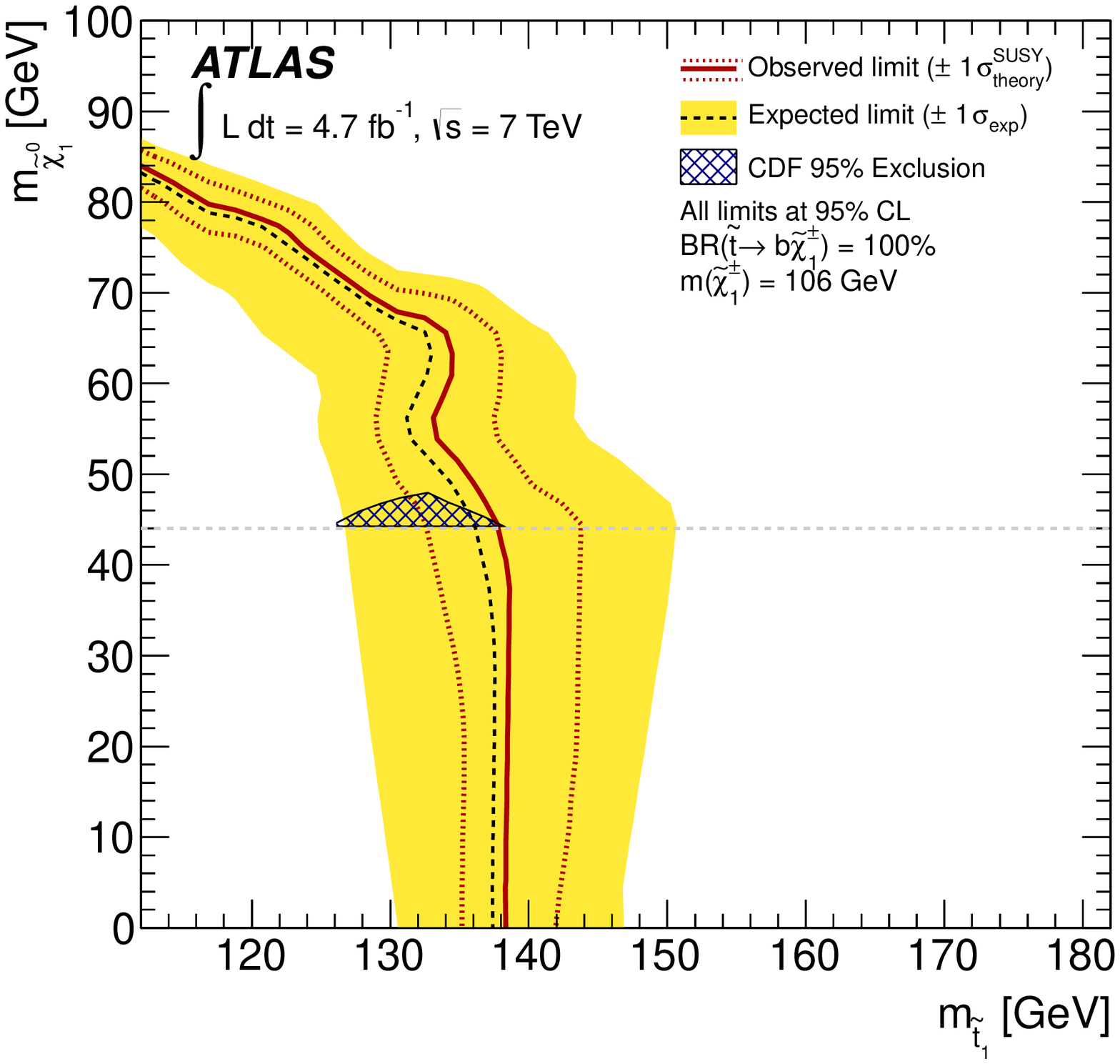}
}
\caption{95\% exclusion limit in the $m_{\tilde{t}_1}$--$m_{\tilde{\chi}_1^0}$ mass plane, with $m_{\tilde{\chi}_1^{\pm}}$=106 GeV. The dashed and solid 
lines show the 95\% C.L. expected and observed limits, respectively, including all
uncertainties except for the theoretical signal cross-section uncertainty (PDF and scale). The band around the expected limit shows the $\pm1\sigma$ result. The dotted $\pm1\sigma$ lines around the observed limit represent the results obtained when moving the nominal signal cross-section up or down by the theoretical uncertainty. Illustrated also is the region excluded at the 95\% C.L. by CDF~\cite{Aaltonen:2009sf}, where the lowest neutralino mass considered was 44 GeV, indicated by the horizontal dotted line.}
\label{fig:limit}
\end{figure}

Table~\ref{tab:results} shows the data observations in the signal regions in each flavour channel, and in the combined flavour channel, along with the evaluated background contributions. Good agreement is observed across all channels, and the absence of evidence for light scalar top production allows a limit to be set on the visible cross-section for non-SM physics, $\sigma_{\rm vis}=\sigma \times \epsilon \times A$, for which this analysis has an efficiency $\epsilon$ and acceptance $A$. The limits are calculated using the modified frequentist CL$_\textup{s}$ prescription~\cite{Read:2002hq} by comparing the number of observed events in data with the SM and SM-plus-signal expectations.

All systematic uncertainties and their correlations are taken into account via nuisance parameters using a profile likelihood technique~\cite{2011EPJC}. In Fig.~\ref{leppt_distributions}, the leading lepton $p_\mathrm{T}$ distributions in the $ee$ and $\mu\mu$ channels are illustrated along with the $E_\mathrm{T}^\mathrm{miss}$ and $E^\mathrm{miss,sig}_\mathrm{T}$ distributions of the data and simulated events in the signal region (with the background normalisations set to their nominal values).

The observed data yield is in good agreement with the SM prediction in the combined flavour channel given in Table~\ref{tab:results}. 

The results in the combined channel are used to place exclusions at 95\% confidence level in the $m_{\tilde{t}_1}$--$m_{\tilde{\chi}_1^0}$ mass plane, using the CL$_\textup{s}$ method. The resulting 95\% confidence level expected (dashed) and observed (solid) limits are shown in Figure~\ref{fig:limit}. Neutralino masses down to 1 GeV are considered, since there is no LEP limit on the neutralino mass in the MSSM for the case that the lightest neutralino is predominantly bino in nature. A bino-dominated lightest neutralino is favoured by the recent LHC Higgs search results.

The observed limits represent a significant extension of the CDF limit~\cite{Aaltonen:2009sf} for a chargino mass of 106~GeV to smaller chargino minus neutralino mass difference (the ATLAS limit extends up to a neutralino mass of 70~GeV for a stop mass of 130~GeV, whilst the CDF limit extends up to a neutralino mass of 46 GeV).

The limit on the stop mass for neutralino masses of 45~GeV (135~GeV) is comparable to the equivalent CDF limit. Increasing the chargino mass by 15~GeV leads to a modest shift of the exclusion limit to higher values of the neutralino mass, with the reach in stop mass being enhanced to a lesser degree due to the falling stop production cross-section. For example, for a model with $m_{\tilde{t}_1}=130$ GeV, $m_{\tilde{\chi}^{\pm}_{1}}=120$ GeV and $m_{\tilde{\chi}^{0}_{1}}=60$ GeV, the value of $A\times\epsilon$ ($0.032\%$) is slightly higher than for the equivalent model with $m_{\tilde{\chi}^{\pm}_{1}}=106$ GeV ($0.026\%$).

\section{Conclusions}
A search for light top squarks has been performed in the dilepton final state. SM backgrounds have been evaluated using a combination of data-driven techniques and MC simulation. Good agreement is observed between data and the SM prediction in all three flavour channels. The results are interpreted in the $m_{\tilde{t}_1}$--$m_{\tilde{\chi}_1^0}$ plane with the chargino mass set to 106 GeV, and with the assumption that the decay $\tilde{t}_1\rightarrow b\tilde{\chi}^\pm_1$ occurs 100\% of the time, followed by decay via a virtual $W$ ($\tilde{\chi}^\pm_1\rightarrow W^{*}\tilde{\chi}^0_1$) with an 11\% branching ratio (per flavour channel) to decay leptonically. A lower limit at 95\% confidence level is set on the stop mass in this plane using the combined flavour channel. This excludes stop masses up to 130~GeV (for neutralino masses between 1~GeV and 70~GeV). This limit exceeds that set by the CDF Collaboration for the same scenario~\cite{Aaltonen:2009sf}.

\section{Acknowledgements}
We thank CERN for the very successful operation of the LHC, as well as the
support staff from our institutions without whom ATLAS could not be
operated efficiently.

We acknowledge the support of ANPCyT, Argentina; YerPhI, Armenia; ARC,
Australia; BMWF, Austria; ANAS, Azerbaijan; SSTC, Belarus; CNPq and FAPESP,
Brazil; NSERC, NRC and CFI, Canada; CERN; CONICYT, Chile; CAS, MOST and NSFC,
China; COLCIENCIAS, Colombia; MSMT CR, MPO CR and VSC CR, Czech Republic;
DNRF, DNSRC and Lundbeck Foundation, Denmark; EPLANET and ERC, European Union;
IN2P3-CNRS, CEA-DSM/IRFU, France; GNSF, Georgia; BMBF, DFG, HGF, MPG and AvH
Foundation, Germany; GSRT, Greece; ISF, MINERVA, GIF, DIP and Benoziyo Center,
Israel; INFN, Italy; MEXT and JSPS, Japan; CNRST, Morocco; FOM and NWO,
Netherlands; RCN, Norway; MNiSW, Poland; GRICES and FCT, Portugal; MERYS
(MECTS), Romania; MES of Russia and ROSATOM, Russian Federation; JINR; MSTD,
Serbia; MSSR, Slovakia; ARRS and MVZT, Slovenia; DST/NRF, South Africa;
MICINN, Spain; SRC and Wallenberg Foundation, Sweden; SER, SNSF and Cantons of
Bern and Geneva, Switzerland; NSC, Taiwan; TAEK, Turkey; STFC, the Royal
Society and Leverhulme Trust, United Kingdom; DOE and NSF, United States of
America.

The crucial computing support from all WLCG partners is acknowledged
gratefully, in particular from CERN and the ATLAS Tier-1 facilities at
TRIUMF (Canada), NDGF (Denmark, Norway, Sweden), CC-IN2P3 (France),
KIT/GridKA (Germany), INFN-CNAF (Italy), NL-T1 (Netherlands), PIC (Spain),
ASGC (Taiwan), RAL (UK) and BNL (USA) and in the Tier-2 facilities
worldwide.

\bibliographystyle{atlasnote}
\bibliography{LightStopPaper}

\onecolumn
\clearpage
\input{atlas_authlist}

\end{document}

%% file: susydefs.tex
\def\lsim{\mathrel{\rlap{\lower4pt\hbox{\hskip1pt$\sim$}}
    \raise1pt\hbox{$<$}}}                
\def\gsim{\mathrel{\rlap{\lower4pt\hbox{\hskip1pt$\sim$}}
    \raise1pt\hbox{$>$}}}                

\makeatletter
\renewcommand{\p@subfigure}{\arabic{figure}}
\renewcommand{\thesubfigure}{\alph{subfigure}}
\renewcommand{\@thesubfigure}{(\thesubfigure)}
\makeatother

%% file: atlas_authlist.tex
\begin{flushleft}
{\Large The ATLAS Collaboration}

\bigskip

G.~Aad$^{\rm 47}$,
T.~Abajyan$^{\rm 20}$,
B.~Abbott$^{\rm 110}$,
J.~Abdallah$^{\rm 11}$,
S.~Abdel~Khalek$^{\rm 114}$,
A.A.~Abdelalim$^{\rm 48}$,
O.~Abdinov$^{\rm 10}$,
R.~Aben$^{\rm 104}$,
B.~Abi$^{\rm 111}$,
M.~Abolins$^{\rm 87}$
% ...
\bigskip

% ...
$^{10}$ Institute of Physics, Azerbaijan Academy of Sciences, Baku, Azerbaijan\\
$^{11}$ Institut de F\'{i}sica d'Altes Energies and Departament de F\'{i}sica de la Universitat Aut\`{o}noma de Barcelona and ICREA, Barcelona, Spain\\
% ...
$^{20}$ Physikalisches Institut, University of Bonn, Bonn, Germany\\
% ...
$^{47}$ Fakult\"{a}t f\"{u}r Mathematik und Physik, Albert-Ludwigs-Universit\"{a}t, Freiburg, Germany\\
$^{48}$ Section de Physique, Universit\'{e} de Gen\`{e}ve, Geneva, Switzerland\\
% ...
$^{87}$ Department of Physics and Astronomy, Michigan State University, East Lansing MI, United States of America\\
% ...
$^{104}$ Nikhef National Institute for Subatomic Physics and University of Amsterdam, Amsterdam, Netherlands\\
% ...
$^{110}$ Homer L. Dodge Department of Physics and Astronomy, University of Oklahoma, Norman OK, United States of America\\
$^{111}$ Department of Physics, Oklahoma State University, Stillwater OK, United States of America\\
% ...
$^{114}$ LAL, Universit\'{e} Paris-Sud and CNRS/IN2P3, Orsay, France\\
% ...
\end{flushleft}

%% file: LightStopPaper.bbl
\providecommand{\href}[2]{#2}\begingroup\raggedright\begin{thebibliography}{10}

\bibitem{Miyazawa:1966}
H.~Miyazawa,
\href{http://dx.doi.org/10.1143/PTP.36.1266}{Prog. Theor. Phys. {\bf 36 (6)}
  (1966)  1266--1276}.

\bibitem{Ramond:1971gb}
P.~Ramond,
\href{http://dx.doi.org/10.1103/PhysRevD.3.2415}{Phys. Rev. {\bf D3} (1971)
  2415--2418}.

\bibitem{Golfand:1971iw}
Y.~A. Golfand and E.~P. Likhtman, JETP Lett. {\bf 13} (1971)  323--326.
[Pisma Zh.Eksp.Teor.Fiz.13:452-455,1971].

\bibitem{Neveu:1971rx}
A.~Neveu and J.~H. Schwarz,
\href{http://dx.doi.org/10.1016/0550-3213(71)90448-2}{Nucl. Phys. {\bf B31}
  (1971)  86--112}.

\bibitem{Neveu:1971iv}
A.~Neveu and J.~H. Schwarz,
\href{http://dx.doi.org/10.1103/PhysRevD.4.1109}{Phys. Rev. {\bf D4} (1971)
  1109--1111}.

\bibitem{Gervais:1971ji}
J.~Gervais and B.~Sakita,
\href{http://dx.doi.org/10.1016/0550-3213(71)90351-8}{Nucl. Phys. {\bf B34}
  (1971)  632--639}.

\bibitem{Volkov:1973ix}
D.~V. Volkov and V.~P. Akulov,
\href{http://dx.doi.org/10.1016/0370-2693(73)90490-5}{Phys. Lett. {\bf B46}
  (1973)  109--110}.

\bibitem{Wess:1973kz}
J.~Wess and B.~Zumino,
\href{http://dx.doi.org/10.1016/0370-2693(74)90578-4}{Phys. Lett. {\bf B49}
  (1974)  52}.

\bibitem{Wess:1974tw}
J.~Wess and B.~Zumino,
\href{http://dx.doi.org/10.1016/0550-3213(74)90355-1}{Nucl. Phys. {\bf B70}
  (1974)  39--50}.

\bibitem{Fayet:1976et}
P.~Fayet,
\href{http://dx.doi.org/10.1016/0370-2693(76)90319-1}{Phys. Lett. {\bf B64}
  (1976)  159}.

\bibitem{Fayet:1977yc}
P.~Fayet,
\href{http://dx.doi.org/10.1016/0370-2693(77)90852-8}{Phys. Lett. {\bf B69}
  (1977)  489}.

\bibitem{Farrar:1978xj}
G.~R. Farrar and P.~Fayet,
\href{http://dx.doi.org/10.1016/0370-2693(78)90858-4}{Phys. Lett. {\bf B76}
  (1978)  575--579}.

\bibitem{Fayet:1979sa}
P.~Fayet,
\href{http://dx.doi.org/10.1016/0370-2693(79)91229-2}{Phys. Lett. {\bf B84}
  (1979)  416}.

\bibitem{Dimopoulos:1981zb}
S.~Dimopoulos and H.~Georgi,
\href{http://dx.doi.org/10.1016/0550-3213(81)90522-8}{Nucl. Phys. {\bf B193}
  (1981)  150}.

\bibitem{lepstop}
{LEP SUSY Working Group (ALEPH, DELPHI, L3, OPAL), Note LEPSUSYWG/01-03.1,
  http://lepsusy.web.cern.ch/lepsusy/Welcome.html}.

\bibitem{Aaltonen:2009sf}
{CDF} Collaboration,
\href{http://dx.doi.org/10.1103/PhysRevLett.104.251801}{Phys. Rev. Lett. {\bf
  104} (2010)  251801}.

\bibitem{Collaboration:2012si}
{ATLAS Collaboration, arXiv:1208.1447 (2012)}.

\bibitem{Collaboration:2012ar}
{ATLAS Collaboration, arXiv:1208.2590 (2012)}.

\bibitem{Aad:2008zzms}
{ATLAS} Collaboration, JINST {\bf 3} (2008)  S08003.

\bibitem{mcatnlolong}
{S.~Frixione, B.~R.~Webber, J. High Energy Phys. 06 (2002) 029; S.~Frixione,
  P.~Nason and B.~R.~Webber, J. High Energy Phys. 08 (2003) 007; S.~Frixione,
  E.~Laenen and P.~Motylinski, J. High Energy Phys. 03 (2006) 092}.

\bibitem{Mangano:2002ea}
M.~L. Mangano, M.~Moretti, F.~Piccinini, R.~Pittau, and A.~D. Polosa,
J. High Energy Phys. {\bf 07} (2003)  001.

\bibitem{herwig}
G.~Corcella et al.,
J. High Energy Phys. {\bf 01} (2001)  010.

\bibitem{Butterworth:1996zw}
J.~Butterworth, J.~Forshaw, and M.~Seymour,
Z. Phys. {\bf C72} (1996)  637--646.

\bibitem{Campbell:1999ah}
J.~M. Campbell and R.~K. Ellis,
\href{http://dx.doi.org/10.1103/PhysRevD.60.113006}{Phys. Rev. {\bf D60} (1999)
   113006}.

\bibitem{Campbell:2011bn}
J.~M. Campbell, R.~K. Ellis, and C.~Williams,
J. High Energy Phys. {\bf 07} (2011)  018.

\bibitem{Aliev:2010zk}
M.~Aliev, H.~Lacker, U.~Langenfeld, S.~Moch, P.~Uwer, et al.,
\href{http://dx.doi.org/10.1016/j.cpc.2010.12.040}{Comput. Phys. Commun. {\bf
  182} (2011)  1034--1046}.

\bibitem{Gavin:2010az}
R.~Gavin, Y.~Li, F.~Petriello, and S.~Quackenbush,
\href{http://dx.doi.org/10.1016/j.cpc.2011.06.008}{Comput. Phys. Commun. {\bf
  182} (2011)  2388--2403}.

\bibitem{powheg}
S.~Frixione, P.~Nason, and C.~Oleari, J. High Energy Phys. {\bf 11} (2007)
  070.

\bibitem{Kersevan:2004yg}
B.~P. Kersevan and E.~Richter-Was, {arXiv:0405247 v1} (2004).

\bibitem{ATLAS:2012al}
{ATLAS Collaboration, arXiv:1203.5015 (2012)}.

\bibitem{Lai:2010vv}
H.-L. Lai, M.~Guzzi, J.~Huston, Z.~Li, P.~M. Nadolsky, et al.,
\href{http://dx.doi.org/10.1103/PhysRevD.82.074024}{Phys. Rev. {\bf D82} (2010)
   074024}.

\bibitem{Sherstnev:2008dm}
A.~Sherstnev and R.~Thorne, {arXiv:0807.2132} (2008).

\bibitem{cteq6l}
{J.~Pumplin et al., J. High Energy Phys. 07 (2002) 012}.

\bibitem{pythia}
T.~Sjostrand, S.~Mrenna, and P.~Skands,
J. High Energy Phys. {\bf 05} (2006)  026.

\bibitem{Beenakker:1997ut}
W.~Beenakker, M.~Kramer, T.~Plehn, M.~Spira, and P.~M. Zerwas,
Nucl. Phys. {\bf B515} (1998)  3--14.

\bibitem{Beenakker:2010nq}
W.~Beenakker, S.~Brensing, M.~Kramer, A.~Kulesza, E.~Laenen, and I.~Niessen,
J. High Energy Phys.. {\bf 1008} (2010)  098.

\bibitem{Beenakker:2011fu}
W.~Beenakker, S.~Brensing, M.~Kramer, A.~Kulesza, E.~Laenen, et al.,
\href{http://dx.doi.org/10.1142/S0217751X11053560}{Int. J. Mod. Phys. {\bf A26}
  (2011)  2637--2664}.

\bibitem{Martin:2009iq}
A.~D. Martin, W.~J. Stirling, R.~S. Thorne, and G.~Watt,
\href{http://dx.doi.org/10.1140/epjc/s10052-009-1072-5}{Eur. Phys. J. {\bf C63}
  (2009)  189--285}.

\bibitem{Botje:2011sn}
{M. Botje, J. Butterworth, A. Cooper-Sarkar, A. de Roeck, J. Feltesse, et al.,
  arXiv:1101.0538 (2011)}.

\bibitem{geant4}
{S.~Agostinelli~et~al.,~Nucl.~Instr.~Meth.~A {\bf 506}~(2003)~250}.

\bibitem{atlassimulation}
{ATLAS Collaboration, Eur. Phys. J. C {\bf 70} (2010) 823}.

\bibitem{Cacciari:2008gp}
M.~Cacciari, G.~P. Salam, and G.~Soyez,
\href{http://dx.doi.org/10.1088/1126-6708/2008/04/063}{J. High Energy Phys.
  {\bf 04} (2008)  063}.

\bibitem{Aad:2011he}
{ATLAS Collaboration, arXiv:1112.6426 (2011)}.

\bibitem{Aad:2011mk}
{{ATLAS}} Collaboration,
Eur. Phys. J. {\bf C72} (2012)  1909.

\bibitem{muons}
ATLAS Collaboration, ATLAS-CONF-2011-063
  \url{https://cdsweb.cern.ch/record/1345743} (2011).

\bibitem{ATLAS-CONF-2011-102}
{ATLAS Collaboration, ATLAS-CONF-2011-102,
  \url{https://cdsweb.cern.ch/record/1369219} (2011)}.

\bibitem{ISRFSR}
{ATLAS Collaboration, Phys. Lett. B 707, 459 (2012)}.

\bibitem{muonefficiency}
{ATLAS Collaboration, ATLAS-CONF-2011-063
  \url{https://cdsweb.cern.ch/record/1345743} (2011)} .

\bibitem{more_unc}
{ATLAS Collaboration, ATLAS-CONF-2011-021
  \url{https://cdsweb.cern.ch/record/1336750} (2011)}.

\bibitem{lumi1}
{ATLAS Collaboration, Eur. Phys. J. C71 (2011) 1630, arXiv:1101.2185}.

\bibitem{lumi2}
{ATLAS Collaboration, ATLAS-CONF-2011-116
  \url{http://cdsweb.cern.ch/record/1376384} (2011)}.

\bibitem{Read:2002hq}
A.~L. Read, \href{http://dx.doi.org/10.1088/0954-3899/28/10/313}{J. Phys. {\bf
  G28} (2002)  2693--2704}.

\bibitem{2011EPJC}
G.~Cowan, K.~Cranmer, E.~Gross, and O.~Vitells, Eur. Phys. J. C {\bf 71} (2011)
   1554.

\end{thebibliography}\endgroup
